# What Obstruct Customer Acceptance of Interne Banking? Security and Privacy, Risk, Trust and Website Usability and the Role of Moderators

**Abstract**

Comparatively a little attention has been paid to the factors that obstruct the acceptance of Internet banking in Sri Lanka. This research assimilates constructs such as security and privacy, perceived trust, perceived risk, and website usability. To test the conceptual model, we collected 186 valid responses from customers who use Internet banking in Sri Lanka. The structural equation modelling technique is applied and hypotheses are validated. The findings show perceived trust and website usability are the possible obstructing factors that highly concerned by Internet banking customers. While security and privacy, and perceived risk are not significant and these are not highly concerned by customers in Internet banking acceptance. The age and gender reveal the moderating effect in each exogenous latent constructs relationship. The practical and managerial implications of the findings are also discussed. This country specific study contributes to the advancement of Internet banking acceptance, and offers some useful insights to researchers, practitioners and policy makers on how to enhance Internet banking acceptance for country similar in context.

## 1. Introduction

The banking sector is one of the fastest growing industries that has adopted Internet banking as a delivery channel for their services (Schierholz and Laukkanen 2007). The advent of the Internet and sophisticated technologies not only stimulated the new industries but it also changed the business model including the banking sector, as a result

Internet banking. It is commonly known as Internet banking or online banking has emerged as a phenomenal growth in recent years (Martins et al. 2014; Riffai et al. 2012; Yang et al. 2015; Yousafzai et al. 2009). It is felt that the growth of banking much depends on the use of Internet banking services rather than traditional banking (Sinha and Mukherjee 2016). Internet banking is the delivery of information or services by a bank to its customers by way of different delivery platforms such as PC banking, Internet banking, managed network and TV-based banking (Daniel 1999). It has many benefits over conventional banking such as 24 hours of service availability, ease of access, elimination of queues, reducing branch operating hours, etc. Hence, Internet banking helps to retain existing customers, advance customer's satisfaction, increase banks' market share, decreasing administrative and operational cost and improve banks' competitive positions (Almogbil 2005; Khalfan et al. 2006). Despite the potential benefits that Internet banking offers to customers, the acceptance of its services have been limited and in many cases, fallen short of expectations (Bielski 2003).

In many developed and developing countries the innovative technologies have accelerated the way banking services are offered, thus consumers being swept along with this trend (Riffai et al. 2012). Prior studies try to explore the Internet banking acceptance in various contexts. For example, trust, usability and perceived quality are believed as the key drivers in Oman for Internet banking acceptance (Riffai et al. 2012). Security/privacy risk is a possible loss due to fraud or a hacker intrude the security of an online bank user (Lee 2009). Service quality, web design and content, security and privacy, convenience and speed where web design and content, convenience and speed are the major factors that influence customer satisfaction in Internet banking (Ling et al. 2016). Trust and distrust are separate constructs and the traditional e-retailing trust had a nomological network and highlighted several distrust experiences whose negative outcome on intention to use online banking (Benamati et al. 2010). Trust is the prominent influencing factor on user's willingness to engage in the online transaction of money and personal sensitive information (Wang et al. 2003). Despite the rising discussion in the literature, yet a limited attempt has been made to study security and privacy, risk, trust and website usability in Internet banking acceptance.

Though prior studies have examined e-banking service acceptance, most of the studies have explored in developed, and only a few study has focused on developing country context. Comparatively, the banking sector in Sri Lanka is young compared to many developed and developing country in the World and Asia region. According to Khan et al. (2017), the Asian countries are becoming more and more involved in Internet banking services, and the customers in these countries are more likely to use e-banking services. In this line of thinking, past studies have addressed the adoption of e-banking from rapidly developing Asian countries' context (Brown et al. 2004; Chan and Lu 2004; Susanto et al. 2013). However, most of these studies are covering only a fraction of e-banking service adoption in the Asian region. In addition, prior studies highlighted that the call for future research to examine Internet banking acceptance either in different countries or other context (e.g., Benamati et al. 2010; Cockrill et al. 2009; Martins et al. 2014; Yousafzai et al. 2009). The recent years' competition, advancement in IT, and customers' lifestyle have changed the face of banking activities among Sri Lankan customers as well. Yet, a very limited attention has been paid on the acceptance of e-banking services in Sri Lanka context. Against this backdrop and from these gap, this study is motivated and try to explore the obstructing factors with moderating role of age and gender. Thus, this study addresses the following research questions:

RQ1. Which factor/s obstruct the acceptance of Interne banking services among banking customers in Sri Lanka?

RQ2. How age and gender have the moderating effect of each obstructing factor – Internet banking acceptance relationship?

In addition, the country-level analysis contributes to the advancement of Internet banking theory and practice by taking national culture into account. As a result, a more rigorous way to see whether the same variables are significant in determining acceptance both across different country context and within different cultures. Hence, this study aims to contribute to the growing literature in Internet banking as Sri Lankan context in the following aspects. First, Sri Lanka could be more viable ground for examining factors motivate customer acceptance of Internet banking, especially after the 30 years of civil war and the financial restoration of its economy. Hence, the country specific research design and finding of this study better fit for other countries that are similar in context. Second, this study integrates moderators into a broader model to investigate the users' acceptance of e-banking services which is lacking in the past studies and highlighted as a limitation in prior studies. Thus, the customer's moderation effect on age and gender will bring more insight about their behavioural pattern in Internet banking acceptance. Third, the theorization and variables used in this study extend the Internet banking acceptance literature. As a result, future studies would integrate this study's model and consider to design more insightful research in this domain.

The remaining of the paper is organized as follows. Section two presents theoretical conceptualization, followed by the research model with hypothesis development. Then, research methodology is presented including instrument development, data collection, and data analysis. Next section presents result and findings. Finally, discussion, implications, limitation and conclusion are presented.

## 2. Theoretical Conceptualization

In recent years, there has been an increasing interest in the technology adoption research. Internet banking services are offered in numerous ways to banking customers. For instance, ATMs became pervasive by the mid-1980s, facilitating banking transactions easier. Telephone banking, firstly human operated and later voice-automated, that reduced the necessity to visit the bank branch. In recent years, the Internet has made banking activities much easier and offer new services to their customers, even eliminating the need to stop at a bank branch. Similarly, Internet banking is expected to become a most important banking method for customers (Nasri and Charfeddine 2012). Though, in reality Internet banking offers many advantages such as faster transaction and lower handling fee, yet there are a large group of customers who refuse to adopt it due to uncertainty and security concern (Chaouali et al. 2016; Lee 2009). Previous study's key attention was largely on single or multiple factors which limit user acceptance of Internet banking services. For example trust (Benamati et al. 2010; Sekhon et al. 2010; Yousafzai et al. 2009), risk (Lee 2009; Martins et al. 2014; Yang et al. 2015), security (Koskosas 2011; Lee and Turban 2001; Polasik and Piotr Wisniewski 2009), privacy (Khalfan and Alshawaf 2004; Lee and Turban 2001), and website usability (Casaló et al. 2008; Pikkarainen et al. 2004). Though, these studies focused the limiting factors for Internet banking acceptance, none of the studies has investigated security and privacy, risk, trust and website usability together with the moderating effect which highly obstructs customer acceptance of Internet banking.

The rapid expansion of Internet technology in recent years has made tremendous changes in the banking sector (Lee 2009; Riffai et al. 2012). As a result, web-based applications are the new ways for banks to retain customers and offer them new services (Martins et al. 2014). Internet banking acceptance has been studied using several models and various conclusions have been drawn from prior studies. Nevertheless, the lack of customer security and privacy, and trust are both in the attribute of the bank and in the Internet banking acceptance have been, and remains barrier in the widespread adoption of Internet banking. While prior research has focused on the factors influencing the e-banking adoption, there is limited empirical research which concurrently addresses acceptance barriers in a wide spectrum. Further, research has shown that in decision making process the schematic processing by women and men is different (Venkatesh and Morris 2000). According to Rodrigues et al. (2016) the behavioral intention depends on the cognitive choice of an online user to respond positively (like) or negatively (dislikes) for online shopping. In addition, in information processing men and women use different socially constructed cognitive structures that, in turn direct them towards perceptions (Venkatesh and Morris 2000). Further, researchers have suggested the inclusion of a set of moderators that remain mostly untested such as culture dimensions (Min et al. 2009) and experience (Venkatesh and Bala 2008). Compare to other possible moderating factors age received less attention as a moderator in prior studies (Sun and Zhang 2006; Wang et al. 2009). Given these significant missing elements, this research paper includes age and gender as the moderators to show the moderating effects for the variable in the model.

In the literature, various theoretical models have been applied to explain the determinants of e-banking adoption among banking customers (Nasri and Charfeddine 2012). The popular theoretical models which describe the relationship of user beliefs, attitudes, and intentions included in theory of reasoned action (TRA), technology acceptance model(TAM), and theory of planned behavior (TPB). TRA suggests that beliefs influence attitudes that in turn direct to intentions and then consequently generate behaviors (Ajzen and Fishbein 1977). Except for the above theories in social psychology, TAM has been validated as an influential theoretical model in explaining IT adoption by users (Davis 1989). Moreover, in 2003 researchers introduced unified theory of acceptance and use of technology (UTAUT) model, which can explain as much as 70% of the variance in intention. The TAM is intended to predict IT acceptance, where perceived usefulness and perceived ease of use are the major determinants of the attitudes which in turn influence behavioral intention to use the actual system (Davis 1989; Venkatesh et al. 2003; Wang et al. 2003). Past research applied TAM model in several countries' context to predict the adoption of Internet banking, example Taiwan (Lee 2009), Tunisia (Nasri and Charfeddine 2012), India (Sinha and Mukherjee 2016), Malay and Chinese ethnic group (Khalil et al. 2010). TAM has been found to have an adequate explanatory power, and the addition of moderators could further improve it (Sun and Zhang 2006). This research model broadens the scope of the adoption decision by focusing four exogenous latent constructs such as security and privacy, perceived trust, perceived risk, and website usability and the role of age and gender as moderators that thoeorized obstructing factors in the accaptence of Internet banking.

Table 01: Prior Studies on Internet Banking

| Study | Research Objective | Sample size & Statistical Test | Study Type | Findings |
|---|---|---|---|---|
| (Chaouali et al. 2016) | Counter-conformity motivation, social influence, & trust in customers' intention to adopt Internet banking. | 245<br>PLS path modelling | Empirical | Trust & customers' counter-conformity motivation influence on performance expectancy. Social influence and trust in the physical bank, have indirect and effort expectancy has no effect |
| (Sinha and Mukherjee 2016) | Investigate why off branch e-banking in India is not accepted as it is in advanced countries | 422<br>multiple regression techniques | Empirical | Trust on technology, trust on bank, perceived ease of use, perceived usefulness, complexity are significantly influence whereas perceived risk was insignificant among customers. |
| (Liao and Wong 2008) | Explores the determinants of customer interactions with e-banking services. | 320<br>Confirmatory factor analysis | Empirical | Perceived usefulness, ease of use, security, convenience & responsiveness significantly explain variation in customer interactions. |
| (Sekhon et al. 2010) | Web site reputation quality of traditional service in a consumer's trustworthiness of e-banking adoption behaviour. | 202<br>Hierarchical moderated regression analysis | Empirical | Service quality makes customer trust and size and reputation of the bank were found to provide structural assurance to the customer. Website features give customers confidence in the e-banking service. |
| (Martins et al. 2014) | Determinants of Internet banking adoption - UTAUT model with perceived risk. | 249<br>Structural equation modelling techniques | Empirical | Performance expectancy, effort expectancy, and social influence, and also the role of risk as a stronger predictor of intention |
| (Riffai et al. 2012) | Customers' acceptance of on-line banking in Oman. | 315<br>Correlation analysis | Empirical | Trust, usability and perceived quality are deemed key drivers and e-banking is skewed to middle aged users. |
| (Yoon and Steege 2013) | Investigate how customers' personalities and perceptions influence Internet banking use. | 125<br>PLS confirmatory analysis | Empirical | Openness, website usability, and perceived security concern significantly influence customers' Internet banking use |
| (Lee 2009) | Predict customers' behavioural intentions in adopting online banking. | 368<br>Confirmatory factor analysis | Empirical | Intention was adversely affected mainly by security/ privacy risk, financial risk is positively affected mainly by perceived benefit, attitude and perceived usefulness. |
| (Roy et al. 2012) | Use TAM model for internet banking adoption in India under security and privacy context. | 619<br>SEM with confirmatory factor analysis | Empirical | Perceived risk has a negative impact on intention of internet banking adoption and trust has a negative impact on perceived risk and well-designed web site found to be helpful in facilitating easier use |
| (Aldas-Manzano et al. 2011) | Role of satisfaction, trust, frequency of use and perceived risk as antecedents of consumer loyalty to banking websites. | 254<br>Confirmatory factor analysis | Empirical | Satisfaction correlates positively with loyalty, the effect is significantly less intense with high levels of perceived risk. trust correlates more positively with high levels of perceived risk, when Internet banking is used less frequently. |

3. Research Model and Hypotheses Development

This study's research model is designed to investigate the factors that obstruct Internet banking acceptance. Hence, this model expands focusing on security and privacy, risk, trust and website usability along with age and gender moderate exogenous and endogenous relationship. The research model is shown in Fig.1 that assimilate together these four construct and hypothesize that they are the key concern by customers during their Internet banking access. The below section discusses the primary motivating literature for each construct with their hypothesis development.

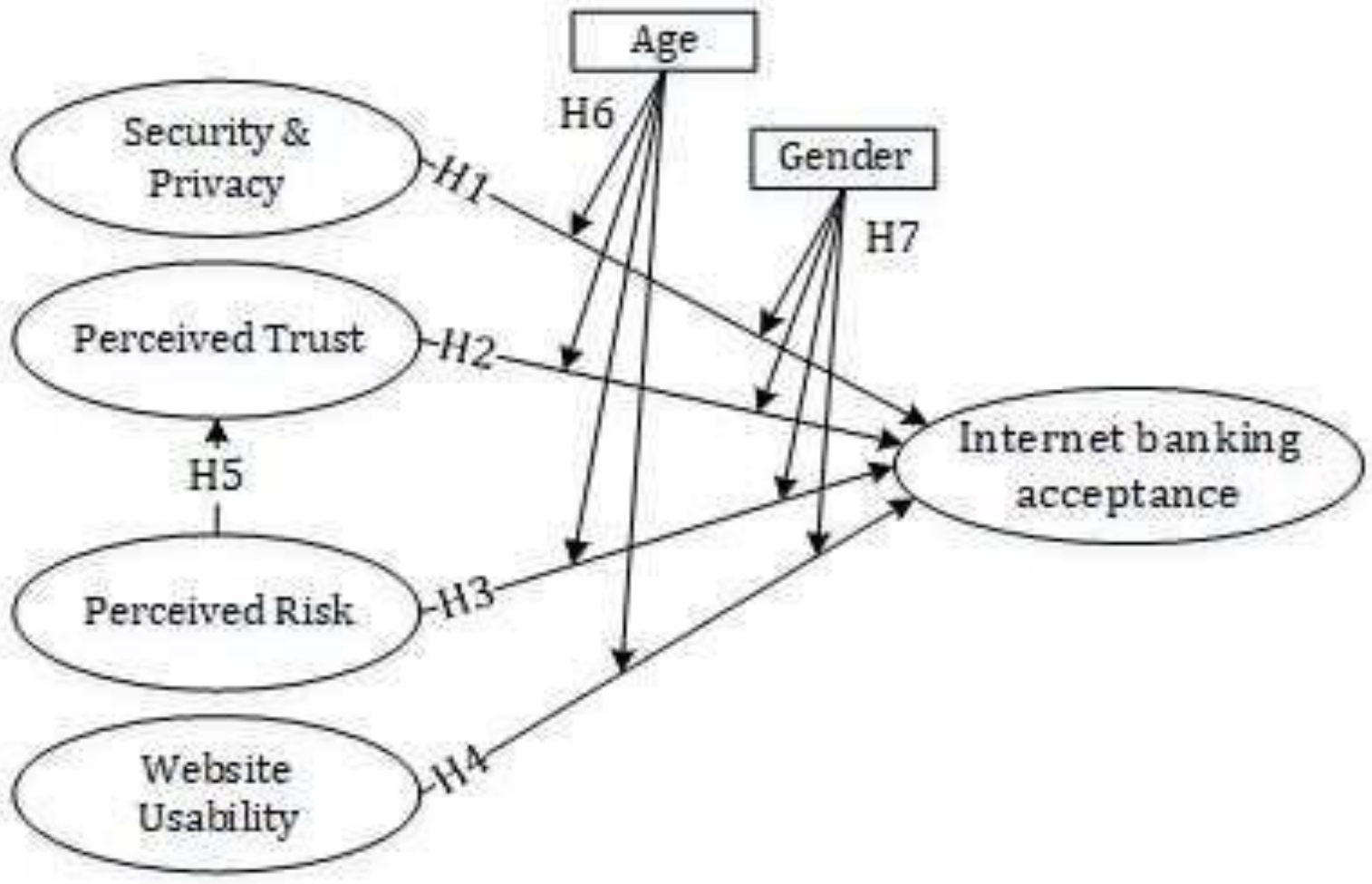

Figure 1: Research Model

3.1. Security and Privacy

Security and privacy in the context of Internet banking defined as "a potential loss due to fraud or a hacker compromising the security of an online bank user" (Lee 2009, p. 2). Internet banking involves financial transaction as primary activities (Liao et al. 2011). Financial transactions using IT devices are tend to the plentiful threat to customers as criminal acts can be done quickly without any physical interaction (Cheung and Lee 2006). As a result, most of the customers are reluctant to adopt Internet banking services due to its security and privacy concerns (Lee 2009; Lee and Turban 2001). The banking industry is associated with high level of trust related to security and privacy issues than traditional banking (Yousafzai et al. 2009). Privacy and security concern are the major sources of dissatisfaction in internet banking (Poon 2007). Today, a number of banking services are offered through the Internet and smart devices, thus customers are highly concerned about security issues more seriously. Privacy and security issues have proven barriers hence, customers keep on eye what kind of data is collected, for what purpose, how long these data is stored, and for what purpose their data is processed (Yoon and Steege 2013). Drawing on the various studies in literature, the issue of security and privacy were found to be the most important determinant factor which obstruct customer adoption of Internet banking (Koskosas 2011; Laforet and Li 2005; Liao and Wong 2008; Polasik and Piotr Wisniewski 2009; Wang et al. 2003; Yoon and Steege 2013). Moreover, information privacy and security issues were identified as serious limiting factors on the adoption and use of e-banking applications (Khalfan and Alshawaf 2004). Accordingly, the security and privacy widely considered as the most important factor which hinders the acceptance of Internet banking among banking customers. Therefore, the first hypothesis is stated as:

H1: The security and privacy are believed as a serious concern among Internet banking customers in Sri Lanka.

### 3.2. Perceived Trust

The trust is defined as "an individual's behavioural reliance on another person under a conditions of risk" (Currall and Judge 1995, p. 3). Trust believed to be a root for adopters' decisions to use new technologies (Gefen et al. 2008). For a customer to have trust in Internet banking, he/she must be made to believe that the transactional medium is secure, and that any information provided to those web sites are not being seized or given to a third party (Suh and Han 2003). In Internet banking, trust is considered as vital aspect because of its "spatial and temporal separation" between the customer and bank (Grabner-Kraeuter 2002) in which, transactions carried out online often do not involve a simultaneous transaction of money. Therefore, in an online mode the absence of direct physical contact creates the nature of service delivery as the lack of trust in Internet banking deter the customer acceptance (Chaouali et al. 2016; Yousafzai et al. 2009). According to Balasubramanian et al. (2003) the customer interactions and perceptions about service attributes such as the reliability of information, availability of the website, confidentiality of information exchange and efficiency of transaction execution makes trust in e-banking environment. On the other hand, no matter consumers are adopters or non-adopters they perceive that they are very much concerned about accessibility and confidentiality in Internet banking (Gerrard and Barton Cunningham 2003). Many studies investigated the influence of trust on consumers' intention to use Internet banking services (Al-Somali et al. 2009; Chaouali et al. 2016; Cockrill et al. 2009; Wang et al. 2003; Yousafzai et al. 2009) and found that trust significantly influences consumers' intention to use Internet banking services. In addition, trust and perceived risk are the direct roots of intention (Gefen et al. 2008; Yousafzai et al. 2009), whereas trust is a multi-dimensional construct with three factors namely; perceived trustworthiness, perceived security, and perceived privacy. The study of Flavian et al. (2006) shows consumer trust and socio demographical traits in which the greater trustworthiness is positively related to higher levels of Internet banking adoption. Moreover, trust is considered as a catalyst in many transactions that can help consumers for satisfying exchange relationships (Hawes et al. 1989). Thus the second hypothesis is stated as follows:

H2: The perceived trust is believed as a serious concern among Internet banking customers in Sri Lanka.

### 3.3. Perceived Risk

Since the 1960s, perceived risk theory has been included to explain consumers' behaviour in the research. Perceived Risk is defined as "the subjectively determined expectation of loss by an online bank user in contemplating a particular online transaction"(Lee 2009, p.2). Whereas Featherman and Pavlou (2003) defined risk as "the potential for loss in the pursuit of the desired outcome of using an e-service." Customers perceive the more money involved in the transaction has the greater the risk. In this case, they will increase their perceived risk and decrease their trust in e-banking (Yang et al. 2015). If a hacker is able to get access to the customer's financial information in the absence of weaker security system which causes for a security related risk create potential loss because of the weakness in the operating system or misuse of funds through criminal access. Hence, perceived risk is posited as a major barrier to consumer acceptance of e-services (Featherman and Pavlou 2003). In this line of thinking, the risk associated with Internet banking is presented from numerous prior studies (Laforet and Li 2005; Safeena et al. 2011). In addition, intention to use online banking is negatively affected mainly by the security/privacy risk, as well as financial risk (Lee 2009). As the impact of risk on the intention of accepting Internet banking is hard to be ignored, the following hypothesis is stated:

H3: The perceived risk is believed as a serious concern among Internet banking customers in Sri Lanka.

### 3.4. Website Usability

According to Barnes and Vidgen (2002) the usability is concerned how a user perceives and interacts with a website, whether is it easy to navigate?, is the design appropriate to the type of the site? and focuses the design principles to improve the website usability. The purpose of aesthetic design is to create an e-banking website visually attractive and enjoyable thus, website design will be a mirror in the customers' satisfaction level (Ahmad and Al-Zu'bi 2011). Website design has been identified one of the determinant factors that influence customer satisfaction not only in e-banking but also in online purchasing site too (Alam et al. 2008; Alam and Yasin 2010; Shergill and Chen 2005). Under website usability Yoon and Steege (2013) includes perceived usefulness (PU) and perceived ease of use (PEU) as these usability factors positively influencing customer's cognition towards Internet banking. In some studies the influence of website usability is addressed such as the information provided in online banking website is the main factor influencing online-banking acceptance in Finland (Pikkarainen et al. 2004). The perceived usability of a website promotes the user's familiarity with the website and it increases the ability to bring website behaviour in the future. Hence, the website usability helps to make information transparent, favours communication and interaction between the parties (Casaló et al. 2008). In literature, website usability has been found and perceived to be the significant factor in determining customer interaction in e-banking (Casaló et al. 2008; Hasbullah et al. 2016; Shergill and Chen 2005; Yoon and Steege 2013). Thus, the hypothesis is stated as follows.

H4: Website usability is believed as a serious concern in Internet banking acceptance among Sri Lankan customers.

## 3.5. Perceived Risk and Trust

The perception of risk is interwoven with trust (Gefen et al. 2008; Liao et al. 2011; Lim 2003). Perceived risk and trust relationship have reached much attention in the literature, but the pragmatic results for the type of relationship are still mixed (Liao et al. 2011; Lim 2003). In Internet banking, the temporal separation between customers and bank and the volatility of Internet services create hidden hesitation on the transaction. Consequently, customers often alert when providing their sensitive and privacy related information and bring their mind towards the role of trust of the transaction channel. Privacy and trust have got much attention in the literature for their relationships with online transactions and also have an effect on one's privacy (Liao et al. 2011). In Internet banking trust will increase the customer likelihood of trusting behaviour as trust is likely to ease concerns of possible negative consequences. Regardless of the increase in Internet banking services, customers still have some reluctance towards them mainly due to the risk concerns and trust related issues (Lee 2009). Prior studies show the possibility for perceived risk to be an antecedent of trust (Dinev and Hart 2006; Liao et al. 2011). Perceived risk is a negative factor that affects customer trust in online transactions (Rouibah et al. 2016). Therefore, the hypothesis is stated as follows:

H5: Customer perceived risk will affect perceived trust among Internet banking customers in Sri Lanka.

Moderators: Age and gender

There is quite a number of factors used as a moderator to study user acceptance of the technology. For instance; Sun and Zhang (2006) classified moderators into three groups such as organizational, technological and individual factors in which age and gender lay in the third group and suggested, research on moderating factors add great value. Prior studies have suggested that gender plays an important role in explaining behavioural intention in information system research (Sun and Zhang 2006; Tarhini et al. 2014; Venkatesh et al. 2003{Tarhini, 2014 #628}). For customer acceptance or rejection choice in the Internet and mobile banking gender and age significantly influenced the decision in Finland (Laukkanen 2016). In this same context sex, income, and age are related to the possibility of e-banking adoption (Flavian et al. 2006). Moreover the study of Yousafzai and Yani-de-Soriano (2012) revealed technology readiness, age and gender moderate the attitude intention among Internet banking users in the UK. In this study we used age and gender as moderators in customer's acceptance determinant factors. Hence, the hypotheses are stated as:

$H_6$: The factors (security and privacy, perceived risk, perceived trust and website usability) on customer acceptance of Internet banking is moderated by age.

$H_7$: The factors (security and privacy, perceived risk, perceived trust and website usability) on customer acceptance of Internet banking is moderated by gender.

## 4. Research Methodology and Data Analysis

### 4.1. Instrument Development

We used the electronic platform to collect the data from Internet banking customers in Sri Lanka. The content validity of the questionnaire was assessed by two senior academics who are expert in empirical research. They checked its accessibility and user friendliness of the electronic version of questionnaire. Their suggestions were incorporated regarding arranging Likert scale question's answering options and rewording the questions with simplified wordings. Questionnaire's reliability was measured by conducting a pilot test sending to 30 banking customers who use Internet banking services. The first author administered the responses from the banking customers. To avoid the multiple responses from banking customers 'limit to one' response per respondent option was set. This research questionnaire includes two sections. The first section covers the demographic profile of the respondent like respondent gender, age, education, occupation, income, types of Internet banking service use and types of IT driven banking service. The second section includes questions to cover the main construct (presented in Appendix C) in the research model with five point Likert scales ranging from ‘‘strongly disagree” =1 to ‘‘strongly agree” = 5.

### 4.2. Data Collection

This study's population is the Internet banking customers including private and state employees, students (undergraduates, postgraduates and others), business people and other e-banking users. Data collection started from June 2015 to May 2016 by sending the electronic version of questionnaire link to 700 banking customers. Due to the less number of responses the link was resent after two months later from the staring date in which users asked to reply if they fail to fill in the first round. The respondent who already answered asked to ignore the second reminder of questionnaire link. Finally, we received 218 responses in total and response rate is 27.95%. Among the received questionnaire, 186 were in a usable condition accounted for 23.85% of valid response rate that is used for the final analysis purpose. The rest of incomplete questionnaires were ignored from the analysis. Table 2 shows the demographic profile of the respondents.

**Table 2: Demographic Profile of the Respondent**

| Gender | *N* | % | Age_Category | *N* | % | Education | *N* | % |
|---|---|---|---|---|---|---|---|---|
| Male | 129 | 69.4 | 18 - 25 years | 52 | 28.0 | GCE O/L | 1 | .5 |
| Female | 57 | 30.6 | 26 - 30 years | 74 | 39.8 | GCE A/L | 7 | 3.8 |
| | | | 31 - 40 years | 38 | 20.4 | Higher Diploma | 21 | 11.3 |
| | | | 41 - 50 years | 18 | 9.7 | Under grad Degree | 98 | 52.7 |
| | | | 51 - 60 years | 4 | 2.2 | Master degree | 53 | 28.5 |
| | | | | | | PhD | 6 | 3.2 |
| | | | | | | | | |
| Income | | | Profession | | | Profession…. | | |
| Below 20,000 Rs | 35 | 18.8 | Student | 28 | 15.1 | Business People | 5 | 2.7 |
| 21,000 - 40,000 Rs | 37 | 19.9 | Teacher | 10 | 5.4 | Foreign employee | 17 | 9.1 |
| 41,000 - 60,000 Rs | 32 | 17.2 | Other state worker | 15 | 8.1 | | | |
| 61,000 - 80,000 | 22 | 11.8 | Academic Staff | 48 | 25.8 | | | |
| 81,000 - 100,000 Rs | 12 | 6.5 | Private Employee | 62 | 33.3 | | | |
| Above 100, 000 Rs | 48 | 25.8 | Self Employment | 1 | .5 | | | |

### 4.3. Data Analysis

The partial least squares (PLS) method of structural equation modelling technique applied to test the model. The usable dataset was electronically imported into SMART PLS 3.0 for the statistical analysis. PLS is mostly used as the statistical technique that has many advantages over other structural equation modelling techniques (Hair Jr et al. 2016; Srivastava and Teo 2007). PLS referred an effective method of analysis for its robustness (Chin et al. 2003; Srivastava and Teo 2007; Subramani 2004). In addition, PLS is recommended for researchers when to measure a larger complex model dealing with attitudes and behaviours (Yoon and Steege 2013).

### 4.4. Measurement Model

To confirm that there is a sufficient discriminant validity among the constructs, the square root of AVE should be greater than the correlations between the construct (Chin et al. 2003; Henseler et al. 2009). As seen in Table 3 for all constructs the square root of the AVE greater than the correlations between the constructs and any other constructs in the model (Hair Jr et al. 2016; Lee 2009). It confirm that all the constructs show evidence for the discriminant validity. In addition, all diagonal values exceeded the inter-construct correlations and it is confirmed that our instrument has suitable construct validity (Hair Jr et al. 2016).

The reliability of individual items, internal consistency between items and the model's convergent and discriminant validity were assessed to ensure the appropriateness of this research measurement model. Table 4 shows items factor loadings, AVE and Cronbach's Alpha for the construct. To analyse the indicator reliability, factor loadings should be statistically significant and preferably greater than the recommended value of 0.7(Chin et al. 2003; Henseler et al. 2009). The t-statistic obtained from bootstrapping (500 subsamples) shows that all loadings are statistically significant at 1%. The construct perceived risk's two items were excluded since they yielded lower factor loading such as PR2 0.554, PR4 0.389 from our initial attempt. After that all items have loadings greater than 0.7 and signify the internal consistency (Hair Jr et al. 2016).

Table 3: Correlation Coefficient and Square Roots of AVEs

| Construct | Mean | SD | Priva_secuty | Perce_Trust | Perce_Risk | Web_Usab | I_bank_accpt |
|---|---|---|---|---|---|---|---|
| Priva_secuty | 3.648 | 0.953 | 0.857 | | | | |
| Perce_Trust | 3.777 | 0.947 | 0.868 | 0.915 | | | |
| Perce_Risk | 3.204 | 1.037 | 0.806 | 0.762 | 0.835 | | |
| Web_Usab | 3.747 | 0.899 | 0.718 | 0.716 | 0.747 | 0.874 | |
| I_bank_accpt | 4.072 | 0.964 | 0.678 | 0.717 | 0.629 | 0.750 | 0.828 |

Note: Diagonal elements are the square root of AVEs, these should exceed the inter construct correlations for adequate discriminant validity. Off-diagonal elements are the correlations among constructs.

For the constructs' reliability, we confirmed two indicators such as composite reliability (CR) and Cronbach's alpha (CA). The most well-known criterion is CA, that provides an estimate of the reliability based on the indicator inter-correlations and assuming that all indicators are equally reliable (Henseler et al. 2009). As seen in Table 4, CR and CA are for each construct above the expected threshold of 0.7, confirming the evidence of internal consistency. To make sure the convergent validity, average variance extracted (AVE) was applied as the AVE is the amount of indicator variance that is accounted for the primary items of construct and should be greater than 0.5, so that the latent variable explains more than half of the variance of its indicators (Henseler et al. 2009). In Table 3, the AVE for each construct is greater than the required threshold of 0.5, confirming the convergent validity (Hair Jr et al. 2016).

**Table 4: Construct, factor loadings, AVE and Cronbach's Alpha**

| Construct | Items | Factor Loadings | t-value | AVE | Composite Reliability | Cronbach's Alpha |
|---|---|---|---|---|---|---|
| Privacy & Security | PS1 | 0.845 | 30.928 | 0.734 | 0.932 | 0.909 |
| | PS2 | 0.902 | 49.963 | | | |
| | PS3 | 0.897 | 45.977 | | | |
| | PS4 | 0.804 | 18.949 | | | |
| | PS5 | 0.831 | 21.306 | | | |
| Perceived Trust | PT1 | 0.913 | 53.033 | 0.838 | 0.963 | 0.952 |
| | PT2 | 0.946 | 84.769 | | | |
| | PT3 | 0.915 | 48.429 | | | |
| | PT4 | 0.894 | 39.227 | | | |
| | PT5 | 0.909 | 40.993 | | | |
| Perceived Risk | PR1 | 0.849 | 28.621 | 0.697 | 0.873 | 0.785 |
| | PR3 | 0.774 | 18.370 | | | |
| | PR5 | 0.877 | 53.348 | | | |
| Website Usability | WU1 | 0.856 | 22.061 | 0.764 | 0.942 | 0.923 |
| | WU2 | 0.882 | 30.248 | | | |
| | WU3 | 0.894 | 42.106 | | | |
| | WU4 | 0.905 | 44.363 | | | |
| | WU5 | 0.832 | 16.449 | | | |
| Internet Bank Acceptance | IBA1 | 0.831 | 23.276 | 0.685 | 0.916 | 0.885 |
| | IBA2 | 0.808 | 16.289 | | | |
| | IBA3 | 0.828 | 23.504 | | | |
| | IBA4 | 0.839 | 23.278 | | | |
| | IBA5 | 0.832 | 19.888 | | | |

# 5. Results and Findings

We employed the two-step procedure suggested by Anderson and Gerbing (1988) for analyzing the data. In the first stage we, checked the measurement model to measure convergent and discriminant validity. Afterwards, we measured the structural model to test the strength and direction of the relations among the theoretical constructs.

### 5.1. Structural model and Hypothesis testing

The structural model can be evaluated by observing path coefficients "þ" that indicates how strong is the relationships between the dependent and independent variables and the ($R^2$) value, which illustrates total variance explained by independent variables (Al-Somali et al. 2009).

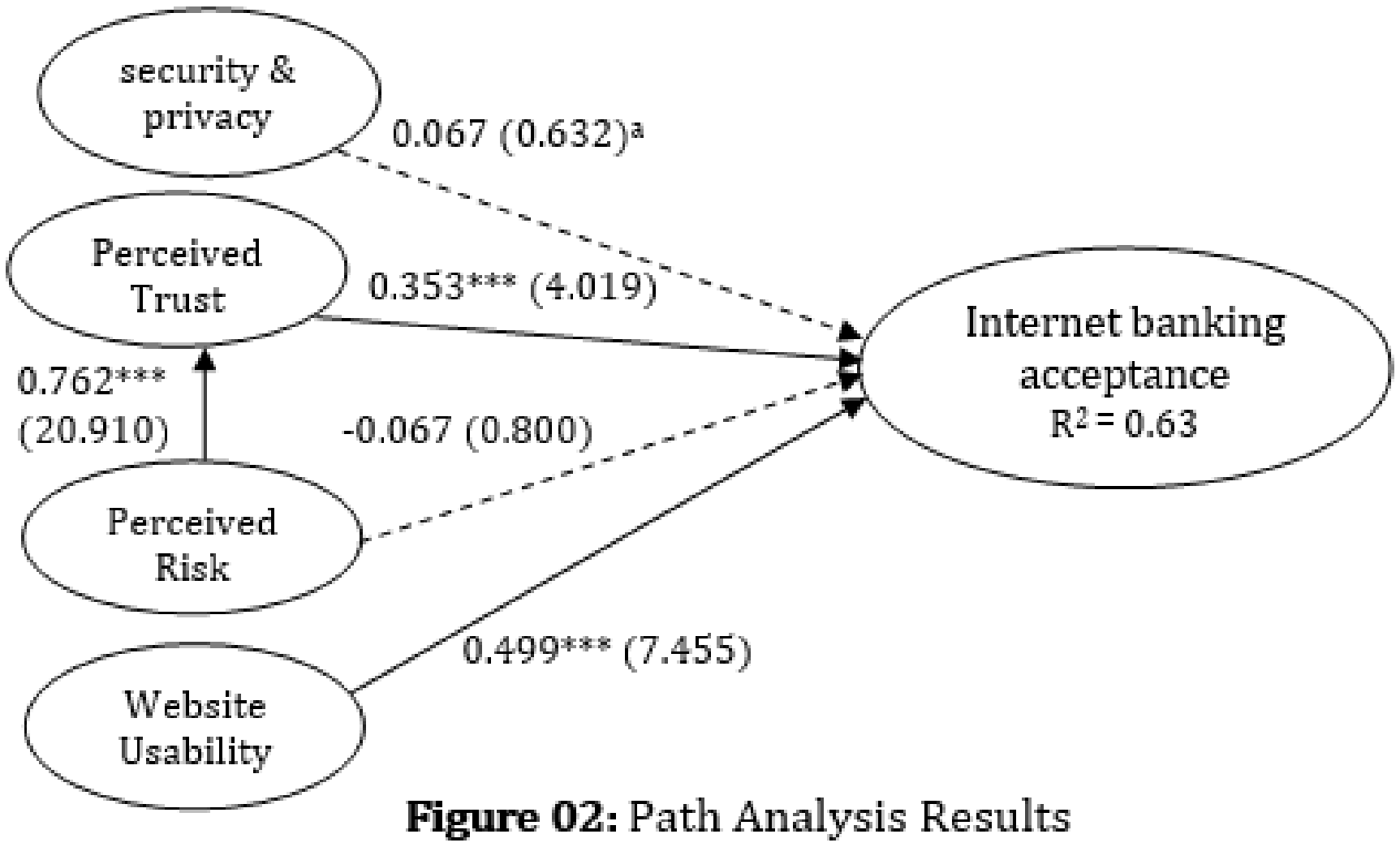

**Figure 02:** Path Analysis Results

Note - [a]: t-values are in bracket ***Significance at $p < 0.001$, **Significance at $p < 0.01$. *Significance at $p <0.05$.

The above Fig 02 shows the final results of the structural model. The path coefficient value with their respective t-value and the non-significant path is shown by dotted lines in Fig 02. The four exogenous latent constructs (security and privacy, perceived trust, perceived risk and website usability) together explained 63% of the variance on Internet baking acceptance. In the above Fig 02 the construct which has positive path effect are such as perceived trust þ = 0.353, website usability þ = 0.499 and Perce_Risk → Perce_Trust þ = 0.762. The security and privacy (þ = 0.067, p= 0.528) and perceived risk (þ = -0.067 p = 0.424) have not shown statistically significant relationship. Thus our hypothesis H2, H4 and H5 are supported and H1 and H3 are not supported. For the statistical significance of parameter estimates, we used t-values (Lee 2009) and p-values. The independence variables' effect is statistically significant such as perceived trust ($t = 4.019$; $p < 0.000$), website usability ($t = 7.455$; $p < 0.000$) and Perce_Risk → Perce_Trust ($t = 20.910$; $p < 0.000$). On the other hand, security and privacy ($t = 0.632$; $p > 0.05$) and Perceived risk ($t = 0.800$; $p > 0.05$) are not statistically significant in Internet banking acceptance. Table 5 shows, the hypothesis testing results of all the variables are given with their structural model assessment.

Table 5: Assessment of the structural model for hypothesis testing

| Hypotheses | Hypothesis Path | Path Coefficient (β) | *t*-Value | p-value | Supported? |
|---|---|---|---|---|---|
| H1 | Priva_secuty → I_bank_accpt | 0.067 | 0.632 | 0.528 | No |
| H2 | Perce_Trust → I_bank_accpt | 0.353*** | 4.019 | 0.000 | Yes |
| H3 | Perce_Risk → I_bank_accpt | -0.067 | 0.800 | 0.424 | No |
| H4 | Web_Usab → I_bank_accpt | 0.499*** | 7.455 | 0.000 | Yes |
| H5 | Perce_Risk → Perce_Trust | 0.762*** | 20.910 | 0.000 | Yes |

Note: ***Significance at $p < 0.001$, **Significance at $p < 0.01$. *Significance at $p < 0.05$.

### 5.2. Summary of Moderating Effect in Each Construct relationship

To examine the moderating effects our sample was categorised gender as male and female, age as younger and older. Accordingly, the sample consists of 129 (69.4%) males and 57 (30.6%) females, whereas age consists of 126 younger (67.7%) and 60 older (32.3%) banking customers. We followed the method applied in the study of Tarhini et al. (2014), in which they split the participant's age to investigate e-learning acceptance in England. We considered initial variable 'age-category' to classify as younger (18 - 30 years old) and older (above 31 years old) in the sample. The table 6 shows the moderating effects of gender and age on the relationship between our four exogenous constructs (security and privacy, perceived trust, perceived risk, and website usability) with dependent variable Internet banking acceptance.

This research employed multi group moderation analysis to assess the moderating effect. The first step, the initial dataset was split into sub datasets based on gender group - male and female, for age group - younger and older. In the second step, the same structural model was run for each dataset separately. In the third step, we followed pair wise comparison of path coefficient and strength of significance difference across the groups considering the p-value. For this purpose, two group's t-statistics were compared with their path difference to assess and find the moderating effect. The t-statistic difference was measured by using the formula suggested by Gaskin (2016) to find moderator variable's significant difference in their respective relationship (see Appendix A). Finally, the conclusions were drawn from the findings.

## 6. Discussion

The research model depicts the relationship of four obstructing factors on Internet banking acceptance. The model consists the possible barriers to accept Internet banking in Sri Lanka and we introduced website usability which is not widely used in customer Internet banking acceptance in the prior studies. As security and privacy have been found to play a significant role in IT adoption research (Khalfan and Alshawaf 2004; Koskosas 2011; Lee and Turban 2001; Liao et al. 2011; Yousafzai et al. 2009) this study also provides evident that security and privacy is a concern in Internet banking use in Sri Lanka. According to Yoon and Steege (2013) to provide stronger theoretical frame the empirical investigation is needed including factors like website quality, security statement, technical protections, etc. This will bring potential outcome in Sri Lankan Internet banking acceptance as well.

This study shows perceived trust is believed as a serious concern among customers in accepting Internet banking. This finding is consistent with the prior studies not only in Internet banking (Chaouali et al. 2016; Cockrill et al. 2009; Flavian et al. 2006), but also studies like online payment as well (Yang et al. 2015). The finding of this research fail not demonstrate a significant relationship between perceived risk and Internet banking acceptance. In general, the higher levels of uncertainty and perceived risk will hinder the acceptance of customers Internet banking adoption. Hence, the role of risk as a stronger predictor of behavioural intention not only in Internet banking acceptance (Martins et al. 2014); but also in online money transaction as well (Liao et al. 2011). Moreover, perceived risk is posited as a significant barrier to consumer acceptance of Internet banking (Lee 2009).

Further, the finding of this research provides strong evidence for website usability which has positive and significant relation with Internet banking acceptance. Thus, Sri Lankan Internet banking customers concern the website usability as a serious concern when they interact with Internet banking. When customers perceive that the website interface has higher quality which tends to have a positive perception of security of the site; this might improve customer intention to use Internet banking (Yoon and Steege 2013). Furthermore, a well-designed web site was also found to be helpful to make easier in Internet banking acceptance (Roy et al. 2012). We tested the interlink of perceived risk and perceived trust. This study also demonstrates that perceived risk concern is strongly and significantly influence on perceived trust. As a result, customers are highly concerned about the seriousness of the risk exists on the Internet banking site and it is believed to be a high concern among Sri Lankan customers. This findings is supported by the prior research such as consumers have built up trust first as an antecedent of their perceived risks and perceived total risk is negatively related to trust (Yang et al. 2015). Further, trust has a negative impact on perceived risk (Roy et al. 2012). Perceived risk decreased customer trust (Rouibah et al. 2016). Trust in the electronic channel and perceived risks of e-commerce are the key determinants of the adoption behaviour (Kim and Prabhakar 2000).

Surprisingly, for gender the findings show that, no strong and significance moderating effect exists among the research variable relationship. But we observed gender has a little moderating effects among the research model constructs. The male group's total variance explained is 69% ($R^2$ = 0.692) and female group's total variance explained is = 51% ($R^2$ = 0.509). In the male group perceived trust has higher difference than female group (t = 3.347). On the other hand, in the female group perceived risk and website usability have higher difference than the male group (Perce_Risk = 1.317, Web_Usab = 5.370). In the case of perceived security approximately male and female groups have equal t-value and a little difference is observed.

Age category explained stronger and significance moderating effects in some exogenous variable relationship. The younger age group's total variance explained is 71% ($R^2$ = 0.710) and the older age group's total variance explained is 46% ($R^2$ = 0.458). In the younger age group perceived trust and website usability have moderating effect than older group (Perce_Trust t = 4.142, Web_Usab t = 5.346) in which perceived trust shows strong and significant moderating effect. On the other hand, in older age group perceived risk and security and privacy have moderating effect than the younger (Perce_Risk t =

1.722, Priva_secuty t = 0.261) where perceived risk has strong and significance moderating relationship. Therefore, our hypotheses H6 and H7 show the moderating effect on the research model and both hypotheses are accepted. Our research findings are consistent with the previous studies, thus the age of the respondent and respondent's gender has a statistically significant impact on Internet banking operations (Polasik and Piotr Wisniewski 2009).

### 6.1. Practical Implication

This research has many practical contributions that are discussed in this section. First, this study finding reveals that perceived trust and website usability are the key concern affecting customer intention to accept Internet banking. Hence, managers should ensure that their Internet banking platform is technically sound, high security practices are needed to minimize the risks, and systems should be up to date. Further, encourage the potential customers by giving information of security and trust on the platform, highly secured channel from customer place to the bank server and handling of sessions with the encryption key. Second, managers can formulate strategies to advertise their service and platform in which their Internet banking is not a risky service, user concerns about computer crimes, invasion of privacy, money back guarantees, and notably display consumer satisfaction guarantees which will make good image and trust about the bank. Third, it is worth noting that banks should focus on the customers who use home personal computer (PC) to access Internet banking. Particularly, younger generation as they are more willing to use it and make them more vigilant. In addition, customers should also be made attentive that banking systems are secured. The security can be guaranteed in Internet banking by showing a privacy statement and displaying trusted third party's logos. For instance, displaying trusted third party logo assures a certain degree of security protection and significantly induce to believe the trustworthiness of bank (Jiang et al. 2008). Fourth, because of the nature of Internet banking which is a less verifiable and controllable environment, one way to improve this security and privacy concern is to offer some form of privacy assurance, such as privacy seal and privacy policy to the customers. The above implications are needed for the bank and managers to retain and expand their current customer base in Sri Lanka.

### 6.2. Limitation and future research direction

This study also has some limitations. First this study did not take into account the social classes based on their income, profession and personality traits. Comparing these classes might show different behavioural pattern towards Internet banking use and which could reveal more significant insight and generalizable findings. Second the sample size for this study is less and more sample will disclose more generalizability of this research findings. Third, it is also more meaningful to include societal and cultural factors. The cultural differences across the country also significant determined with the different cultural background in the region among South Asian countries context will give more generalizable findings. Fourth the most of the respondents are young and well knowledgeable customer whose behavioural intention might be different from the average customers. These customers are likely willing to accept and more familiar in accessing Internet banking. Therefore, this may have biased our findings. Fifth, factors like social awareness, lifelong experiences, changes on user's privacy, security and trust concern which may take place at different time interval. Finally, for this study primary data is used to draw the conclusion. If we could use longitudinal data by observing customer behavioural changes and concern in different period will give more insight into the phenomenon. Future research can take into account these limitations and presents the potential outcome it has.

## 6.3. Conclusion

Despite the growing importance of Internet banking acceptance, to our knowledge very limited studies analyse key factors which obstruct the customer acceptance of Internet banking in Sri Lanka. In overall, the model developed in this research has practical implications as it helps to identify factors which barriers for the acceptance of Internet banking. One of the key contributions in this study is the inclusion of age and gender in the model which are highlighted in the prior study as a limitation (Yoon and Steege 2013). It is also noteworthy that our findings show perceived trust and website usability are the possible obstructing factors which are highly concerned by Sri Lankan Internet banking customers. The constructs security and privacy, and perceived risk fail to confirm that they are the serious concern among customers in accepting Internet banking. This may be due to the security and privacy, trust, and website usability issues of the users' cognitive intention may have prejudiced toward this construct. This study's moderators (age and gender) revealed that they have an effect on each exogenous and endogenous relationship. Moreover, perceived privacy and perceived risk also show the causal relationship in our model. Drawing upon this research findings, this country specific study is the good direction which can be applicable for countries which are similar in context.

**Appendix:A**

$$t = \frac{Path_{sample_1} - Path_{sample_2}}{\left[\sqrt{\frac{(m-1)^2}{(m+n-2)} * S.E._{sample1}^2 + \frac{(n-1)^2}{(m+n-2)} * S.E._{sample2}^2}\right] * \left[\sqrt{\frac{1}{m} + \frac{1}{n}}\right]}$$

**Appendix** : B Moderating Effect Comparison of Age and Gender

| Constructs | Male $R^2$ = 0.692 (129) | | Female $R^2$ = 0.509 (57) | | Statistical Comparison of path | Younger $R^2$ =0.710 (126) | | Older $R^2$ = 0.458(60) | | Statistical Comparison of path |
|---|---|---|---|---|---|---|---|---|---|---|
| | Standardized path coefficient | t-value | Standardized path coefficient | t-value | | Standardized path coefficient | t-value | Standardized path coefficient | t-value | |
| Perce_Risk -> e-bank_adop | -0.011 | 0.198 | 0.145 | 1.317 | t = 1.023 P=0.308 | -0.072 | 0.809 | 0.257 | 1.722* | t = 2.063 ** P=0.041 |
| Perce_Trust -> e-bank_adop | 0.429 | 3.347 *** | 0.185 | 1.603 | t = 1.187 P=0.237 | 0.464 | 4.142*** | 0.118 | 0.990 | t = 1.907 * P=0.058 |
| Priva_secuty -> e-bank_adop | 0.024 | 0.189 | 0.019 | 0.152 | t = 0.025 P=0.980 | 0.004 | 0.085 | -0.028 | 0.261 | t = 0.163 P=0.871 |
| Web_Usab -> e-bank_adop | 0.436 | 4.565 *** | 0.466 | 5.370*** | t = 0.192 P=0.848 | 0.496 | 5.346*** | 0.398 | 3.523*** | t = 0.619 P=0.536 |

Note: ***Significance at $p < 0.01$. **Significance at $p <0.05$ *Significance at $p < 0.10$.

**Appendix** : C Construct development methodology

| Construct Type | Sub Construct | Source |
|---|---|---|
| Internet banking acceptance | IBA1 I am using e-banking services as it fulfills my banking needs.<br>IBA2 I intend to use e-banking if the cost & times reasonable to me<br>IBA3 In future, I plan to use e-banking services very often<br>IBA4 I am using e-banking services as it has convenient, security, trust and easiness.<br>IBA5 I intend to increase my usage of the e-banking services. | (Al-Somali et al. 2009; Foon and Fah 2011; Khalil et al. 2010) |
| Perceived Trust | PT1 My e-banking website is secure and trustworthy<br>PT2 I trust the transaction conducted via e-banking is secure<br>PT3 I trust payments made via e-banking channel will be processed securely and accurately<br>PT4 I believe my personal information on e-banking will be kept confidential<br>PT5 I can rely on e-banking services to use as I expected | (Al-Somali et al. 2009; Khalil et al. 2010; Yee-Loong Chong et al. 2010), |
| Privacy and Security | PS1 E-banking does not misuse my personal information<br>PS2 I feel secure in providing sensitive information for my e-banking transactions<br>PS3 E-banking makes me feel safe with my online transactions<br>PS4 I feel privacy and security procedure are vital in e-banking service | (Liao and Wong 2008; Yousafzai et al. 2009) |
| | PS5 I am satisfied with e-banking security system which is available in Sri Lanka | |
| Perceived Risk | PR1 I feel free to submit my personal and confidential information via e-banking transaction<br>PR2 When transaction error occurs; I worry that I cannot get compensation from bank<br>PR3 Existing government policies are sufficient to keep e-banking transactions<br>PR4 I am worried to use e-banking services because others may able to access my account<br>PR5 I have confidence in the security of existing e-banking service network | (Yousafzai et al. 2009) |
| Website Usability | WU1 I can easily navigate through the web content and web pages<br>WU2 The structure and contents of this e-banking website is clear and understandable<br>WU3 The experience that I had with this e-banking website was satisfactory<br>WU4 When navigating this e-banking web site , I feel I am in control of what I can do<br>WU5 The organization of the contents, design and user friendliness make it easy for me to know where I am when navigating it. | (Barnes and Vidgen 2002; Jaruwachirathanakul and Fink 2005) |

**Appendix: D Item to construct loadings**

| | Inte_Bank_Accept | Per_Risk | Per_Trust | Secr_Priv | Web_Usab |
|---|---|---|---|---|---|
| IBA1 | 0.831 | 0.538 | 0.624 | 0.633 | 0.695 |
| IBA2 | 0.808 | 0.458 | 0.561 | 0.472 | 0.618 |
| IBA3 | 0.828 | 0.515 | 0.582 | 0.566 | 0.609 |
| IBA4 | 0.839 | 0.580 | 0.632 | 0.606 | 0.633 |
| IBA5 | 0.832 | 0.501 | 0.560 | 0.510 | 0.529 |
| PR1 | 0.505 | 0.849 | 0.673 | 0.703 | 0.587 |
| PR3 | 0.417 | 0.774 | 0.464 | 0.492 | 0.544 |
| PR5 | 0.621 | 0.877 | 0.729 | 0.777 | 0.719 |
| PT1 | 0.688 | 0.732 | 0.913 | 0.813 | 0.708 |
| PT2 | 0.662 | 0.702 | 0.946 | 0.814 | 0.687 |
| PT3 | 0.644 | 0.694 | 0.915 | 0.773 | 0.614 |
| PT4 | 0.595 | 0.667 | 0.894 | 0.790 | 0.573 |
| PT5 | 0.687 | 0.692 | 0.909 | 0.781 | 0.687 |
| PS1 | 0.583 | 0.652 | 0.722 | 0.845 | 0.578 |
| PS2 | 0.573 | 0.723 | 0.769 | 0.902 | 0.623 |
| PS3 | 0.594 | 0.735 | 0.802 | 0.897 | 0.625 |
| PS4 | 0.564 | 0.613 | 0.715 | 0.804 | 0.572 |
| PS5 | 0.587 | 0.724 | 0.706 | 0.831 | 0.674 |
| WU1 | 0.622 | 0.609 | 0.603 | 0.635 | 0.856 |
| WU2 | 0.687 | 0.669 | 0.651 | 0.665 | 0.882 |
| WU3 | 0.655 | 0.706 | 0.678 | 0.658 | 0.894 |
| WU4 | 0.669 | 0.661 | 0.633 | 0.627 | 0.905 |
| WU5 | 0.641 | 0.616 | 0.561 | 0.552 | 0.832 |